\documentclass[preprint,12pt]{article}

\usepackage{amssymb}
\usepackage{amsmath}
\usepackage{doi}
\usepackage{graphicx}
\usepackage[giveninits=true]{biblatex}
\usepackage{hyperref}
\usepackage{url}

\title{Topological characterisation of a chaotic attractor with an additional branch generated from economic data}

\author{
    \begin{minipage}{\textwidth}
        \centering
    Alexandre Meneceur\textsuperscript{1},
    Vincent Lignon\textsuperscript{2},
Martin Rosalie\textsuperscript{1}\\
\bigskip
{\small
1. Laboratoire Génome et Développement des Plantes, Université de Perpignan Via
Domitia, CNRS, Avenue P. Alduy, Perpignan, 66860, France \\
2. Centre de Droit Économique et du Développement Yves Serra, Université de Perpignan
Via Domitia, Avenue P. Alduy, Perpignan, 66860, France 
}
\end{minipage}
}

\begin{document}

\maketitle

%% Abstract
\begin{abstract}
There are insights of chaotic properties in economic systems and data. 
To prove the existence of chaotic dynamics, the establishment of a deterministic model is mandatory.
A global modelling tool (GPoM) is used to search for mathematical models of equations from economic data: unemployment, inflation and nominal exchange rate over 30 years.
A system of three differential equations is chosen as a model, whose solution is a chaotic attractor in $\mathbb{R}^{3}$.
The model extracted from the data is not able to fit them, but it provides equations linking those multiple economic variables and reveals significant impact of exchange rate on unemployment and inflation evolution.
The topological characterisation of the chaotic attractor solution exhibits an additional branch in its first return map to the Poincaré section. 
Consequences of this particular structure are analysed and interpreted economically.
\end{abstract}


\newpage
\section{Introduction}

Chaotic dynamics are often considered as a dead end when they are observed in an experimental setup or in time series.
This is because, contrary to standard dynamics that are characterised by stabilisation or periodic oscillations, chaotic dynamics remain unpredictable even if they are solutions of deterministic systems.
This deterministic property is mandatory to prove the existence of chaos, and a mathematical model ensures this \cite{Glass_2009}.

There exist tools that compute values giving insight on the fact that chaos could appear in economic systems \cite{Olmedo_2011, Faggini_2014}. 
Indeed, economic fluctuations are made of complex systems with multiple interactions where chaotic dynamics might arise.
As a consequence, the investigation of chaotic dynamics within economic systems has emerged as a significant area of research \cite{baumolbenhabib, day}: this exploration is motivated by the limitations of linear models in capturing the complexities and nonlinearities inherent to economic phenomena.

The methodologies employed in economics can be broadly categorised into two distinct approaches. The foremost and significantly most prevalent method \emph{(the theoretical approach)} involves the construction of theoretical economic models that, through specific parametrisations or structural designs, exhibit chaotic behaviour. It often begins with established economic frameworks, such as  general equilibrium models and introduces nonlinear elements or mechanisms that can generate complex dynamics \cite{Bischi_2024}. 

The second one, which is more investigative and less developed \emph{(the empirical approach)}, focuses on the empirical analysis of economic data to detect the presence and characteristics of chaotic dynamics: it seeks to identify chaotic patterns directly from observed economic data, such as stock prices, exchange rates, or macroeconomic indicators. Based on a single time series, Elena Olmedo  \cite{Olmedo_2011} has for instance underlined that unemployment in Spain behave in a chaotic way.

The purpose of the paper is to provide a topological characterisation and interpretation of a chaotic attractor generated from economic data.
We go further in the empirical approach of chaos in economics using French multiple times series data.
The aim is to extract structure and dynamical properties from time series to have insight on the mechanisms relating multiples economic variables: forecasting and data fitting is not our goal.

From this perspective, we focus on the relationship between inflation and unemployment, which is central to macroeconomic analysis -through the Phillips curve- and provides a rich ground for exploring complex dynamics \cite{phillips1958}. The Phillips curve, in its original form, posits an inverse relationship between inflation and unemployment, suggesting a trade-off that policymakers can exploit. However, the stability and predictability of this relationship have been questioned over time, with empirical evidence suggesting that the Phillips curve can shift or even disappear under certain conditions \cite{Lavoie01102024, gordon2011history}. Recently, François Geerolf \cite{geerolf} argues that the conventional understanding of this curve may not accurately reflect the complexities of modern economies. More specifically, he suggests that a third component -the exchange rates- plays a significant role in the economic dynamics that link unemployment and inflation.
 
In this framework, we propose to use tools such as GPoM \cite{Mangiarotti2012} to generate systems of differential equations from three economic time series (inflation, unemployment and the nominal exchange rate). As the tools generate polynomial and non-linearities, it could provide a chaotic attractor as solution (e.g. \cite{mangiarotti2019chaos}).
Currently, several tools can help to describe, characterise and classify these chaotic dynamics \cite{Letellier_2021}.
The solution of the obtained system exhibits a particular structure with a unimodal first return map to the Poincaré section with an additional branch.
Such a particular structure can be interpreted in multiple ways as a prolongation in the template \cite{Aguirre_2008, Abadie_2025} or as an extra branch \cite{Letellier_2001, Kamdoum_Tamba_2016}.
Theoretical results presented hereinafter using the Rössler system \cite{rossler1976equation} will clarify the role of such an additional branch.
Finally, the topological analysis of the chaotic attractor solution of a model will provide some details on the dynamics and its implication on the economical trajectories and interpretations.

\section{Topological characterisation method}
\label{sec:topological_characterisation}

Further to this recent review \cite{Letellier_2021}, chaos is defined as follows: ``A more practical definition could be that a solution $S$ to a dynamical system $f$ : $\mathbb{R}^d \to \mathbb{R}^d$ is said to be chaotic if $f$ is deterministic, $S$ is bounded, and $S$ is sensitive to initial conditions.''
Consequently, the deterministic aspect of models is fundamental to obtain chaos and, the differential equations systems obtained from global modelling techniques ensure this property.
Literature on \emph{chaotic dynamics} and \emph{nonlinear dynamics} has been developed at the end of the century while the main concepts of this theory come from the early nineties (see Fig.~1 of \cite{aguirre2009modeling}).

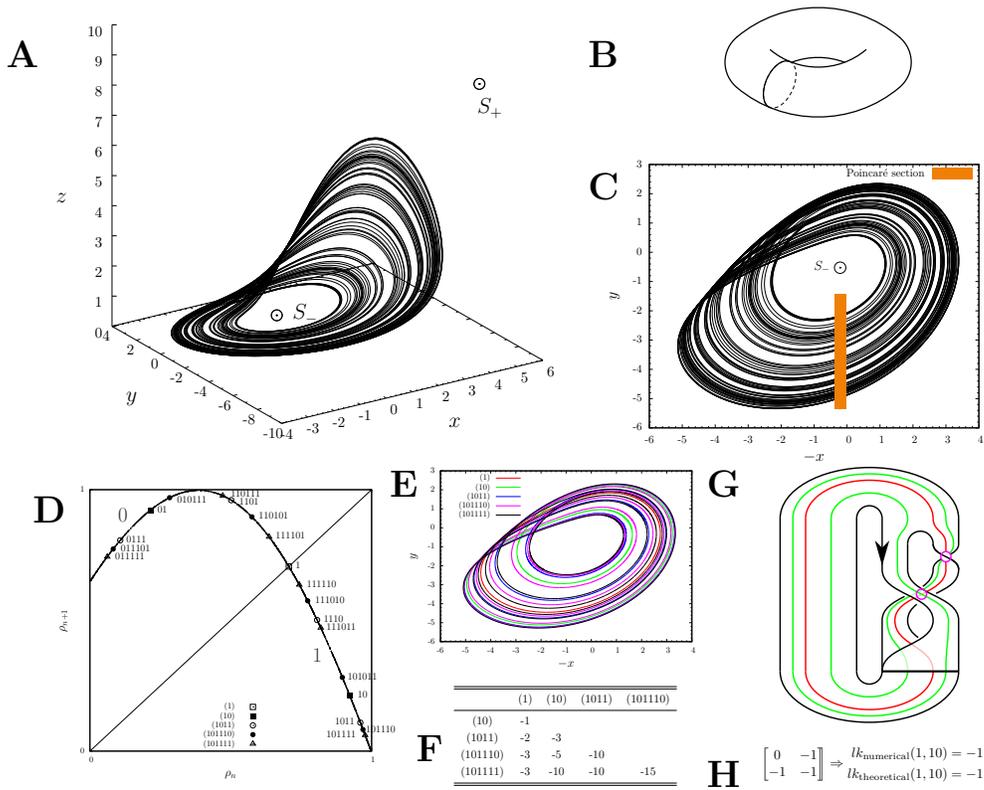
\begin{figure}[htpb]
    \centering
    \includegraphics[width = .9\textwidth]{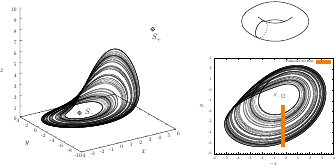}
    \put(-370,150){\large \textbf{A}}
    \put(-150,150){\large \textbf{B}}
    \put(-150,100){\large \textbf{C}}
    
    \includegraphics[width = .9\textwidth]{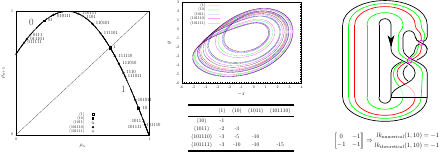}
    \put(-360,100){\large \textbf{D}}
    \put(-225,110){\large \textbf{E}}
    \put(-215,10){\large \textbf{F}}
    \put(-105,110){\large \textbf{G}}
    \put(-105,0){\large \textbf{H}}
    \caption{
        \textbf{Topological characterisation of a chaotic attractor.}
        A. $\mathcal{R}$ is the chaotic attractor solution of the Rössler system \eqref{eq:rossler}.
        B. Bounding torus of the attractor indication the required number of components to create the appropriate Poincaré section.
        C. Position of the Poincaré section \eqref{eq:rossler_01_section_rho} in the phase space with a flow evolving clockwise.
        D. First return map with periodic points indicating the position of the Unstable Periodic Orbits (UPOs).
        E. Numerical representation of the UPOs.
        F. Table containing numerical linking numbers between UPOs.
        G. Template describing the topological properties of the attractor.
        H. Linking matrix detailing torsions and permutations for each branch in the template with the validation when theoretical linking numbers between orbits correspond to numerical linking numbers.
    }
    \label{fig:rossler}
\end{figure}

Figure~\ref{fig:rossler} represents the steps of the topological characterisation of a chaotic attractor solution to a standard system in the domain \cite{Letellier_1995}. The Rössler system \cite{rossler1976equation}:
\begin{equation}
  \left\{
    \begin{aligned}
      \dot{x} &= -y -z \\
      \dot{y} &= x +ay \\
      \dot{z} &= b+z(x-c)
  \end{aligned}
  \right.
  \label{eq:rossler}
\end{equation}
has the attractor $\mathcal{R}$ (Fig.~\ref{fig:rossler}A) solution for parameters $a = 0.398$, $b= 2$ and $c=4$.

The main tools developed for analysing attractors solution of dissipative systems are the Lyapunov Exponents (quantification of the rate of separation for infinitesimally closed trajectories), and the Poincaré section (Figure~\ref{fig:rossler}b) whose purpose is to transform a continuous problem to a discrete representation of trajectories and their link after a topological period.
For the Rössler system, the Poincaré section is defined as follows:
\begin{equation}
  \mathcal{P} = \left\{ (y_n,-z_n) \in \mathbb{R}^2\ |
  -x_n = -x_-, \ -\dot{x}_n<0\right\} 
  \label{eq:rossler_01_section_rho}
\end{equation}
where $x_-$ is a $x$-value of the central fixed point of the system.
From this Poincaré section, a variable $\rho_n$ is defined to represent the distance from the inside of the attractor to the outside:
\begin{equation}
    \rho_n = 1 - (y_n + 4.98)/(-2.15+4.98)
\end{equation}
This variable is used to build the \textit{first return map} that is the signature of the chaotic dynamics (Fig.~\ref{fig:rossler}D).
Unstable Periodic Orbits (UPOs) could be considered as the skeleton of a chaotic attractor because of their unstable property (Fig.~\ref{fig:rossler}E) \cite{Letellier_1995}.
Consider that a differential equations system is numerically solved (using the four-order Runge-Kutta algorithm, RK4), the solution is a trajectory in a phase space.
Thus, at a moment, this numerical solution of a differential equations system will ``follow'' one UPO for a while before diverging and visiting another UPO.
This process maintains the trajectory into the attractor; the latter being the solution of the differential equations system that cannot described analytically.
The ``periodic'' in periodic orbits refers to the state space and not to the time space (topological period).
Periodic orbits are time invariant while the system evolves in a chaotic state (from initial condition, the solution evolves and successively visits the unstable
periodic orbits).
From a chaotic time series that is the numerical solution of a differential equations system, UPOs can be extracted from the first return map, and this acquisition is a preliminary step of the \textit{topological characterisation} \cite{gilmore1998topological}.
For dissipative systems, the purpose of this method is to obtain the structure of the chaotic mechanism from a topological invariant (the linking number Fig.~\ref{fig:rossler}F) computed between every couple of UPOs (the reader is referred to \cite{gilmore2002topology} for details).
The template was originally introduced as ``knot-holder'' by Birman and Williams \cite{Birman_1983} that permits to resume the topological properties of the topological invariants extracted from the attractor using the linking number (Fig.~\ref{fig:rossler}G).
The template is a formal description of the branched manifold detailing the topological properties of a chaotic attractor.
The linking numbers between each pair of orbits are computed numerically and theoretically using the algebraic description of the template $R$ \eqref{eq:rossler_template}.
The template is validated if both are equals for corresponding couple of UPO (Fig.~\ref{fig:rossler}G, H).
A template is described with a linking matrix detailing the number of branches with their torsions and how they permute before joining.
The template of the Rössler attractor is thus defined by the linking matrix:
\begin{equation}
    R = \left.\left[ \begin{matrix}
    0 & -1 \\ -1 & -1
    \end{matrix} \right]\!\!\right].
    \label{eq:rossler_template}
\end{equation}
This formalism underlines the chaotic mechanism of the attractor with a splitting chart that separate the flow in branches (sensitivity to initial conditions with the stretching and folding part of the mechanism) and a joining chart where trajectories are regrouped before going for another round (bounded solution).
Templates of various chaotic attractors solution to a Rössler system have been established using this framework \cite{Rosalie2016b}.

\section{Modelling economic data and chaotic solution}
\label{sec:economic_model}

\subsection{Global modelling}
\label{ssec:global_modelling}

The global modelling technique initiated in the early 1990s \cite{crutchfield1987equations, AGUIRRE_1995, Gouesbet_1994} aims to extract differential equations systems directly from observational time series.
Originally, it was mostly applied to single time series and to either theoretical \cite{Gouesbet_1991, AGUIRRE_1995} or experimental cases \cite{Letellier_1995b}.
Applications to multivariate time series from real environmental conditions is recent \cite{Mangiarotti_2015}.
Few tools have been developed for this purpose.
The Generalized Global Polynomial Modelling (GPoM) package \cite{Mangiarotti2012, Mangiarotti_2019} developed at Cesbio by S. Mangiarotti is one of them.
Another popular technique, the Sparse Identification of Nonlinear Dynamics (SINDy) tool \cite{Brunton_2016}, was proven to be quicker, but its ability to obtain non trivial approximations of environmental dynamics remains uncertain.
From this point of view, the GPoM platform has proven to be particularly interesting.
It was successfully applied to numerous realms, which makes it the best candidate for the present project.
Since the early 2010s, it has been successfully applied to agronomy of cereal crops cycles \cite{mangiarotti2014two}, geohydrology of karstic systems \cite{mangiarotti2019chaos}, eco-epidemiology of plague \cite{Mangiarotti_2015},
epidemiology of Ebola and COVID-19 \cite{mangiarotti2020chaos}, eco-hydrology of earthworms activity \cite{mangiarotti2021earthworm}, and seasonal cycles of CO$_2$ in caves \cite{saez2024scenarios}.

GPoM can be used with different objectives: (1) for dynamical characterisation (chaos detection and topological classification); (2) for coupling detection (in particular under highly nonlinear conditions), and (3) for retro-modelling (that is, to obtain interpretable equations). Note that for this latter objective, in a single case, a complete interpretation of the differential equations systems obtained from observational time series could be proposed, \cite{Mangiarotti_2015} which reveals the potential of the approach.
The GPoM package is an open-source package developed in R language and made available on the Comprehensive R Archive Network (CRAN) \cite{GPoM}.

\subsection{Economic data, pre-treatment and origin}

\begin{figure}[htbp]
    \centering
    \includegraphics[width=.9\textwidth]{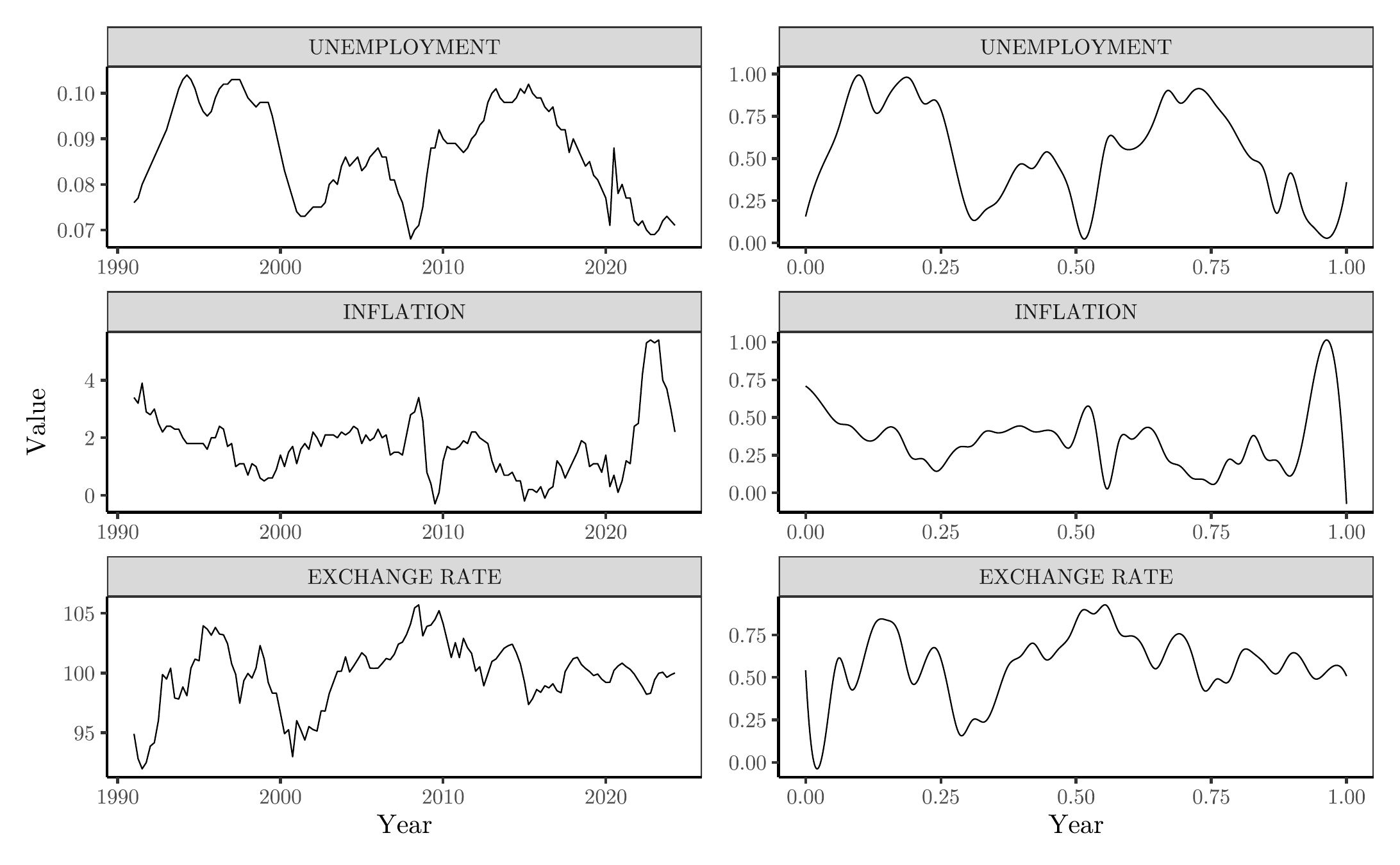}
    \caption{Original time series (first column) and the normalized and resampled time series (second column) used to perform global modelling with GPoM.}
    \label{fig:plotvs}
\end{figure}

A set of three-monthly time series is considered; each associated to a macroeconomic variable in the French setting. 
Those variables have been chosen accordingly to the recent debates on the Phillips curve mentioned above \cite{geerolf}. 

The time series on unemployment and inflation come from the INSEE~\cite{insee} database, whereas the time series on exchange rate comes from the BIS~\cite{BIS} database over 30 years. 
Fig.~\ref{fig:plotvs} represents the time series of the three variables:
\begin{itemize}
    \item
    Unemployment rate (the share of the labour force that is unemployed and actively seeking employment)
    \item
    Inflation rate (the percentage increase or decrease in the price level of goods and services in the economy over a year)
    \item
    Nominal effective exchange rate (a measure of the value of a country's currency relative to a weighted basket of foreign currencies)
\end{itemize}
According to the GPoM documentation, it is often necessary to subsample the time series, before resampling them using spline interpolation, as giving a polynomial form to the curves may help the algorithm succeed in finding a nonlinear dynamic. 
The values of the time series are also normalised in $\left[0, 1\right]$ to prevent the presence of enormous coefficients in the system (see second column of Fig.~\ref{fig:plotvs}).
The GPoM algorithm is then applied to the set of time series, with different parameters to ensure that most combinations of polynomial terms are covered by the algorithm.
Numerous models are therefore produced by the algorithm, and we focus on choosing the one that embodies the most the dynamics of the original time series.

\subsection{Model}

\begin{figure}[htb]
    \centering
    \includegraphics[width = .5\textwidth, trim = {80 80 80 80}, clip]{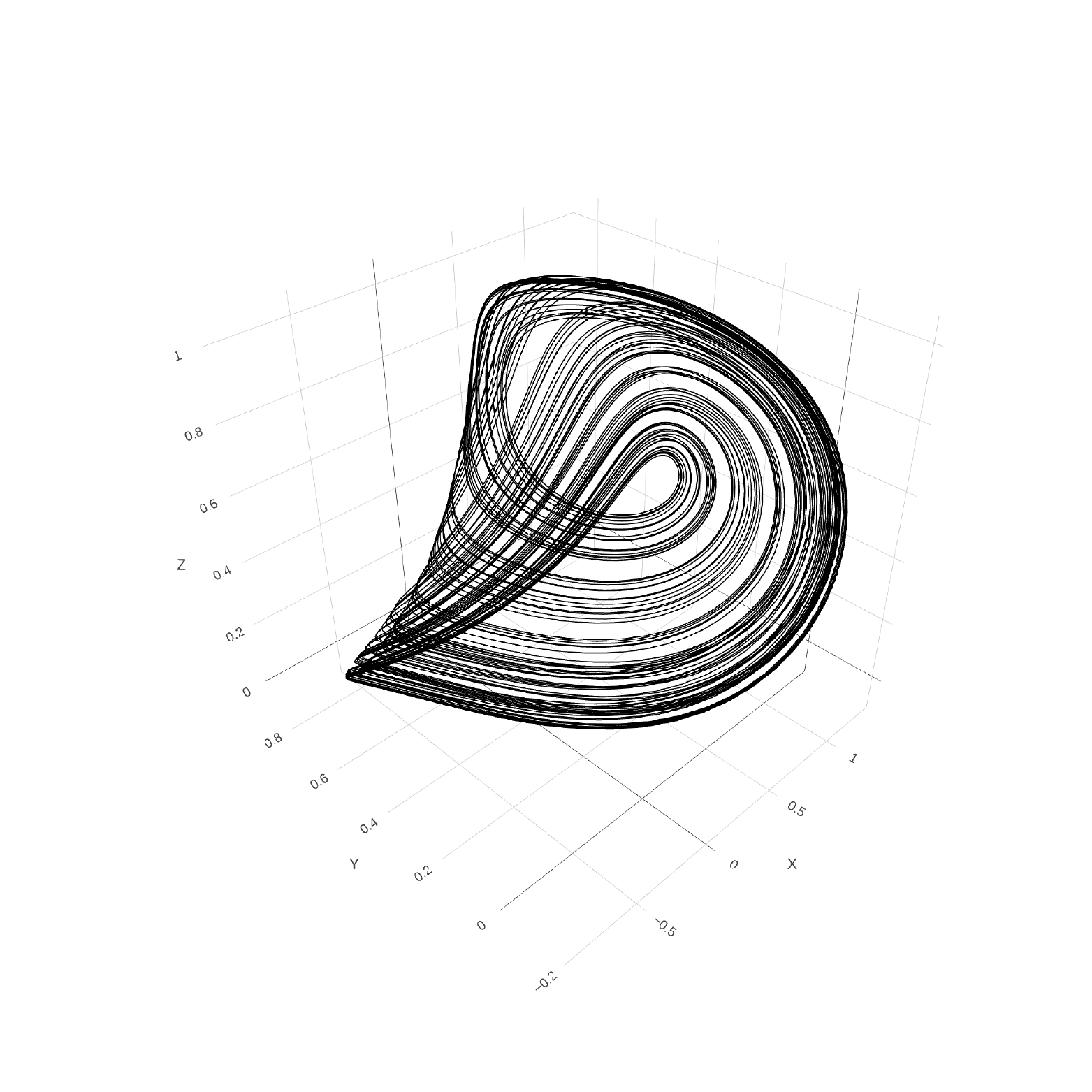}
    % \put(-210,160){\large \textbf{A}}
    % \quad
    % \includegraphics[width = .45\textwidth]{PlotAttracteur.pdf}
    % \put(-220,190){\large \textbf{B}}
    \caption{
    \textbf{Attractor $\mathcal{E}$ solution of the model \eqref{eq:economic}.}
    % A. Three-dimensional representation of the chaotic attractor solution of \eqref{eq:economic}. % B. The $(x,y)$ plane with equilibrium points. 
    }
    \label{fig:economic_attra}
\end{figure}

\begin{figure}[htb]
    \centering
    \includegraphics[width = .5\textwidth]{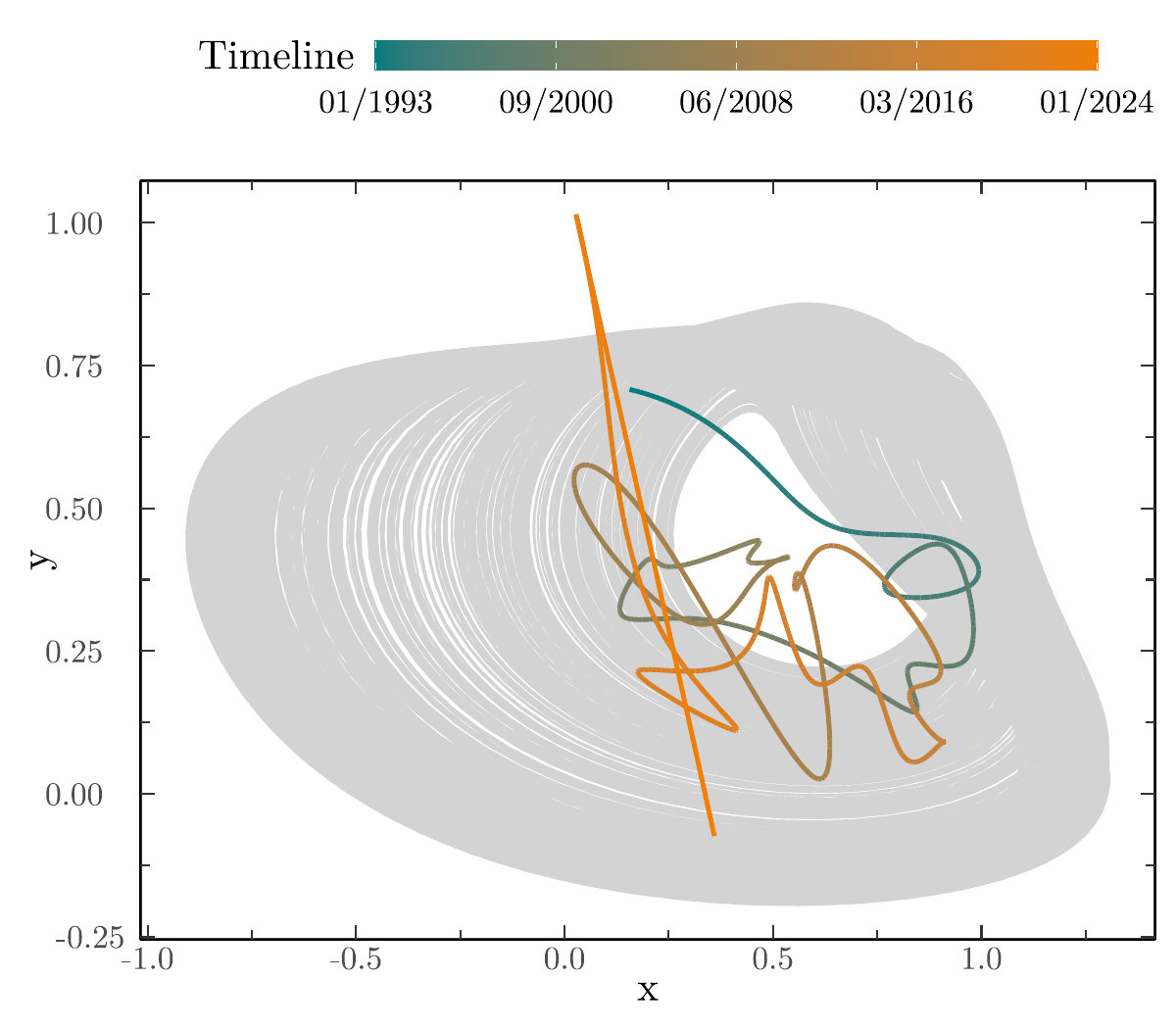}
    \caption{
        \textbf{Differential equations system and the original values.}
        $x$ corresponds to the unemployment rate and $y$ to the inflation rate.
        The original time series projected in the $(x, y)$ phase space mainly evolves clockwise.
        The solution of the model \eqref{eq:economic} in grey also evolves clockwise.
    }
    \label{fig:serie_plus_attra}
\end{figure}

For the reminder of this article, the following notation is used:
\begin{itemize}
    \item
    Unemployment rate (x)
    \item
    Inflation rate (y)
    \item
    Nominal effective exchange rate (z)
\end{itemize}
and the chosen model generated by GPoM is defined as follows:
\begin{equation}
    \left\{
    \begin{aligned}
        \dot{x} &=-15.2193492215949  + 12.6246703650997\ z  + 12.2891205846936\ z^2  \\ &\quad + 38.2077775976987\ y -51.327806215602\ yz 
        \\
        \dot{y} &= 11.3194672143246\ z -11.1810798067421\ z^2 -4.59572630441015\ x 
        \\    
        \dot{z} &=-5.1609358283762\ xz  + 10.5277054934486\ xy 
    \end{aligned} \right.
    % \caption{ODE system outputted by GPoM}
    \label{eq:economic}
\end{equation}

First, it indicates that the data could be connected using these relations.
The solution of this system is a chaotic attractor $\mathcal{E}$.
Fig.~\ref{fig:serie_plus_attra} shows the projection of the time series on the attractor that is structured even if the dynamics is chaotic.
As our purpose is not to fit the trajectories, this mathematical model is a way to reveal the complex underlying dynamics contained in the data.

Then, the model outputted by GPoM is partly consistent with the theoretical framework proposed by the Phillips Curve and its developments \cite{geerolf}. 
It reflects the original intuition that the unemployment rate has a negative impact on inflation, which is consistent with the idea that an improvement in the situation on the labour market maintains upward pressure on prices (through higher household consumption and greater bargaining power of employees). 
Besides, the nominal exchange rate has a positive (and cumulative) impact on the unemployment rate. 
Indeed, an appreciation of the national currency (i.e. the exchange rate) can lead to a loss of competitiveness for the country (the price of exported products being more expensive for non-resident economic agents). 
In this context, when the currency appreciates, export-oriented industries may experience a drop-in activity and lay off part of their workforce.

More generally, this model suggests that the nominal exchange rate could play a role in the cross-relation between inflation and unemployment (insofar as the exchange rate depends on the central bank's monetary policy reactions). For instance, the evolution of the unemployment rate depends on the level of inflation, but the sign of this effect is conditional on the level of the nominal exchange rate. In other words, there could be one or more nominal exchange rate thresholds that could have a strong impact on the combined dynamics of unemployment and inflation. 

The relationships highlighted by the model are not theoretically supported by structural approaches in economics. While this exploratory model is not intended to replace them, it does suggest that cross-relationships involving the exchange rate (and therefore central bank intervention) can lead to very different dynamics that could be taken into account in theoretical modelling \cite{soliman}.

\section{First return map with additional branches}
\label{sec:first_return_map}

\subsection{First steps of the topological characterisation}

\begin{figure}[btbp]
    \centering
    \includegraphics[width = .45\textwidth,trim= 25em 25em 25em 25em,clip]{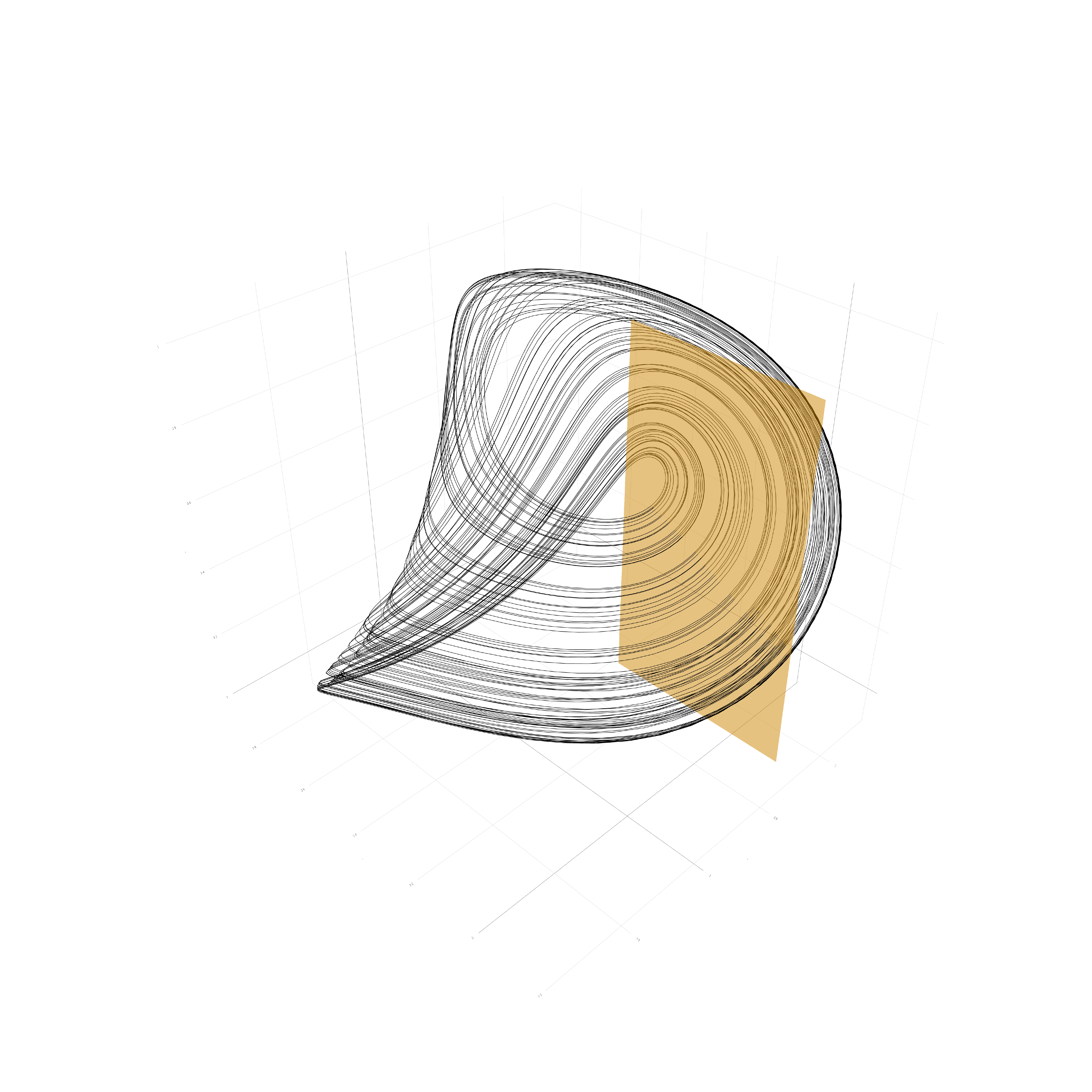}
    \put(-170,160){\large \textbf{A}}
    \quad
    \includegraphics[width = .45\textwidth]{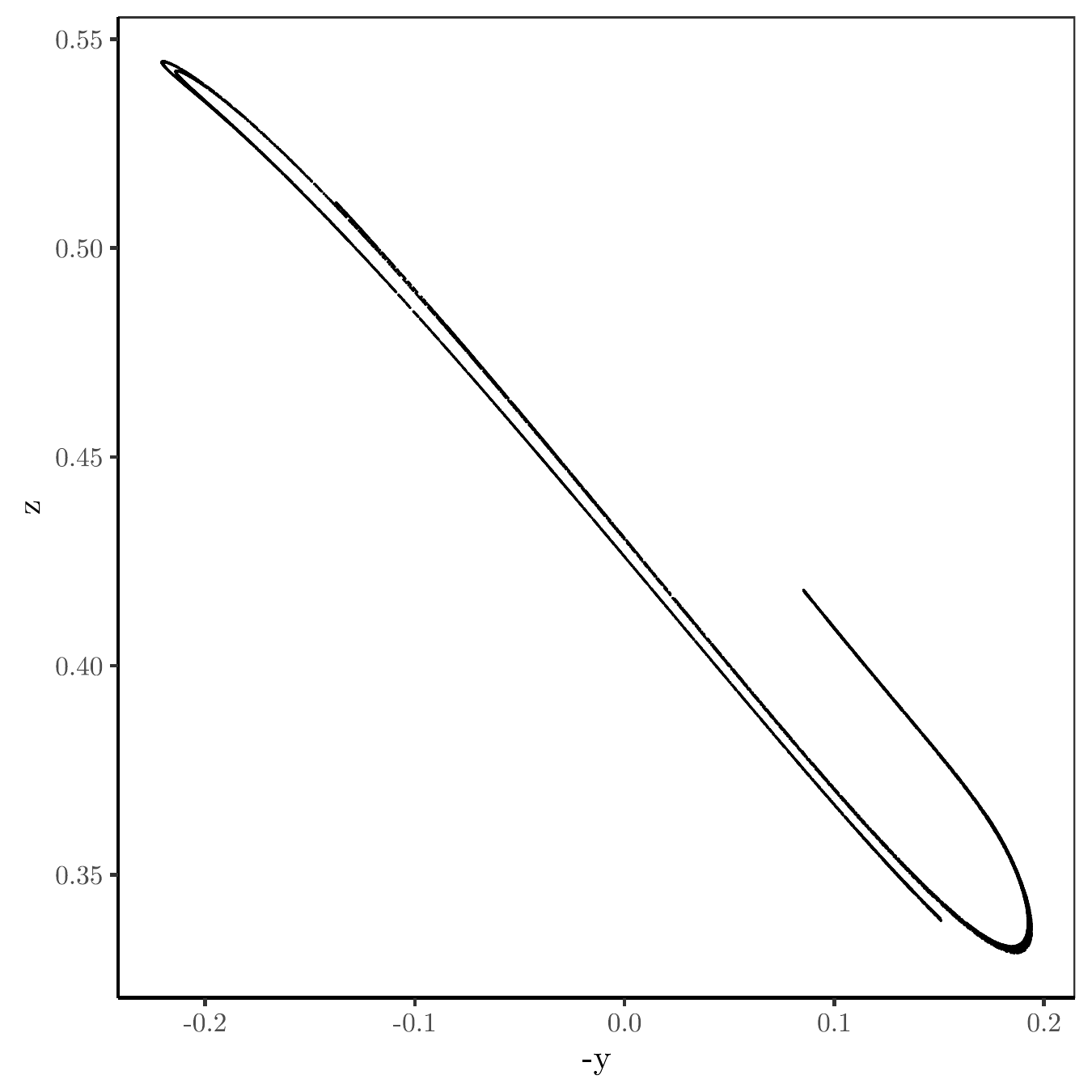}
    \put(-180,160){\large \textbf{B}}
    \caption{A. Three-dimensional representation of the chaotic attractor solution $\mathcal{E}$ and the Poincaré section $\mathcal{P}$. B. Computed points of $\mathcal{P}$.}
    \label{fig:poincaresec}
\end{figure}

The topological analysis is performed as previously defined (Sec. \ref{sec:topological_characterisation}). 
The flow evolves clockwise in the phase plane $(x-y)$.
The Poincaré section is the hyperplane $\mathcal{P}$ (represented in Fig.~\ref{fig:poincaresec}.A) defined as follows:
\begin{equation*}
    \mathcal{P}= \left\{ (y(t),z(t)) \in \mathbb{R}^2\ | x(t) =  x_-,\ y(t) < y_-,\ t > 0\right\}
\end{equation*}
Fig.~\ref{fig:poincaresec}B shows the Poincaré section with the orientation of the $y$--axis from the inside to the outside of $\mathcal{E}$.
This Poincaré section exhibits a particular structure where there are several layers that corresponds to chaotic dynamics signature where the flow is stretched, folded and squeezed.
There is no possibility to position a Poincaré section on $\mathcal{E}$ without having such a structure with multiple layers.
To precisely understand the implication of such a constraint on topological characterisation, we first address this problem with the well known Rössler system \cite{rossler1976equation}.

\subsection{The case of the Rössler attractor}
\label{ssec:rossler}

\begin{figure}
    \centering
    \includegraphics[width = .45\textwidth]{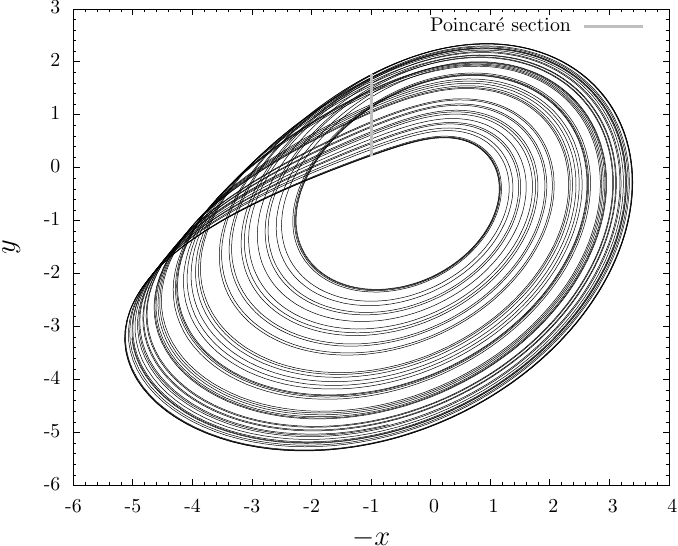}
    \put(-180,130){\large \textbf{A}}
    \quad
    \includegraphics[width = .45\textwidth]{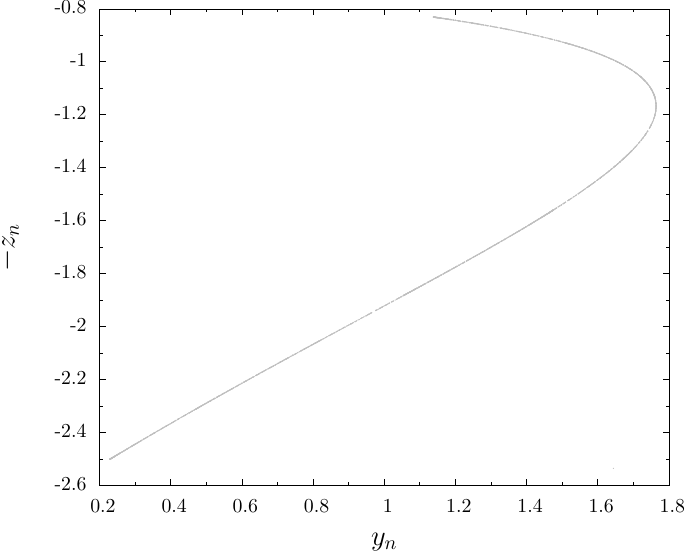}
    \put(-180,130){\large \textbf{B}}
    \caption{
        \textbf{Poincaré section positioned inside the chaotic mechanism.}
        A. Space phase $(-x, y)$ with flow evolving clockwise with the Poincaré section $\mathcal{Q}$ \eqref{eq:rossler_01b_section_gamma} place in the stretching and folding mechanism.
        B. Poincaré section $\mathcal{Q}$ \eqref{eq:rossler_01b_section_gamma} in $(y_n, -z_n)$. The only use of $y_n$ to build a variable $\gamma_n$ remove the bijection between the section and the Poincaré section.
    }
    \label{fig:rossler_01b}
\end{figure}

\begin{figure}
    \centering
    \includegraphics[width = .45\textwidth]{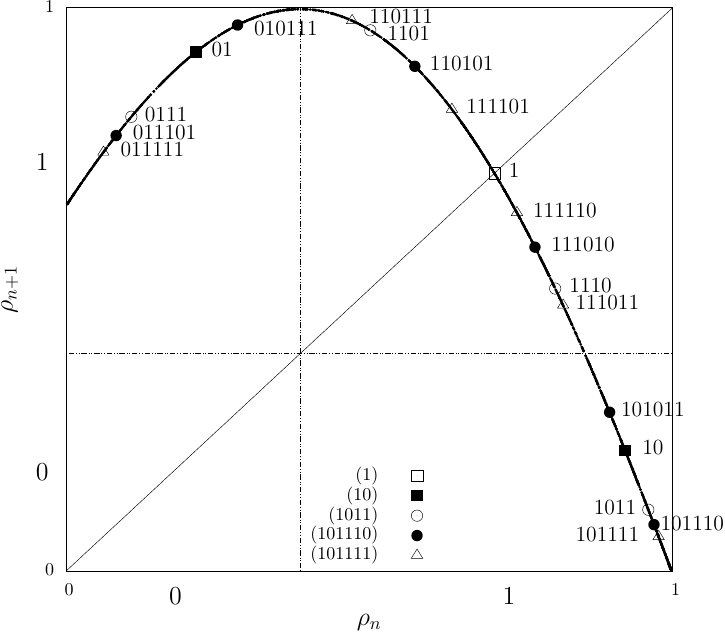}
    \put(-180,140){\large \textbf{A}}
    \includegraphics[width = .45\textwidth]{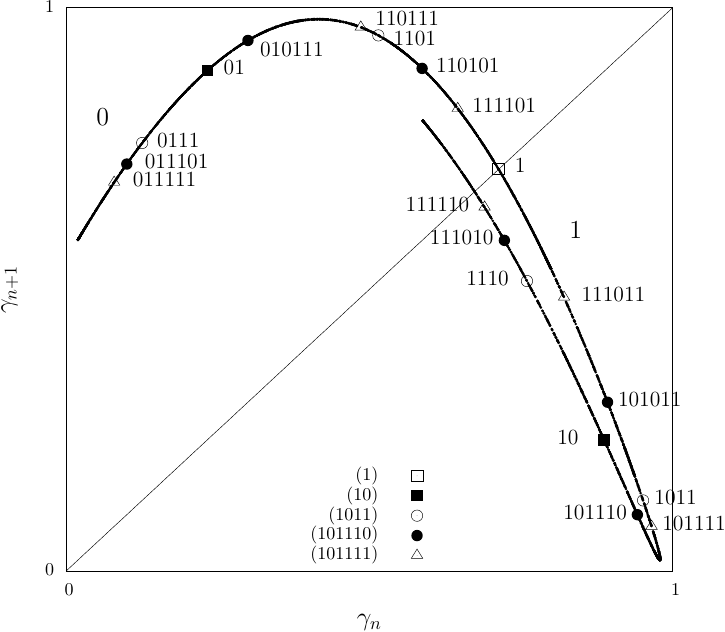}
    \put(-180,140){\large \textbf{B}}
    \caption{
    \textbf{Comparison of periodic points order in first return maps.}
    First return maps of two equivalents Poincaré sections for the Rössler system with periodic points of the unstable periodic orbits. A and B correspond respectively to Poincaré section \eqref{eq:rossler_01_section_rho} and \eqref{eq:rossler_01b_section_gamma}. As these two first return maps are obtained from equivalent Poincaré sections, the relative order of the periodic points is conserved by following the line in A and B. This translates the fact that the stretching and folding mechanism start in the three-dimensional space inducing modification in the structure of the first return map B compared to A. In addition, the relative order of periodic points corresponds to $\rho_n$ values in A while it does not correspond to the order of $\gamma_n$ values in B.} 
    \label{fig:rossler_01ab_appli_orbites}
\end{figure}

\begin{figure}[htb]
    \centering
    \includegraphics[width = .45\textwidth]{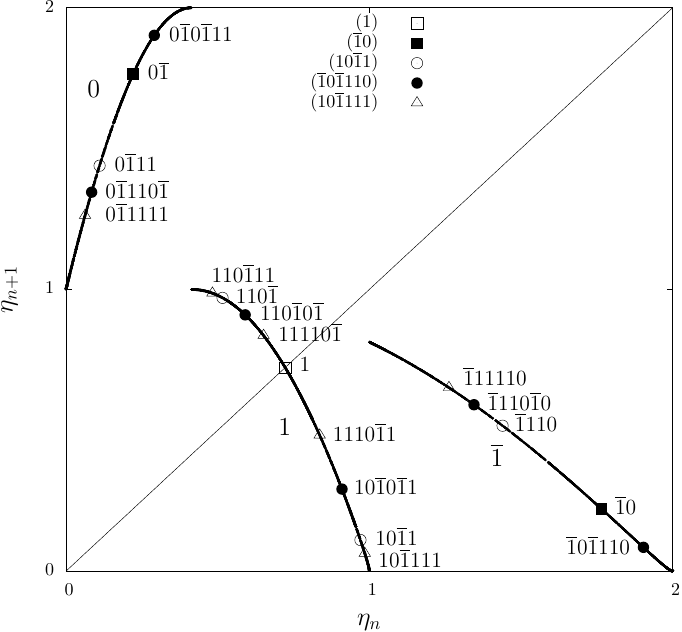}
    \put(-182,150){\large \textbf{A}}
    \qquad \qquad \includegraphics[width = .3\textwidth, trim=0 5cm 0 0]{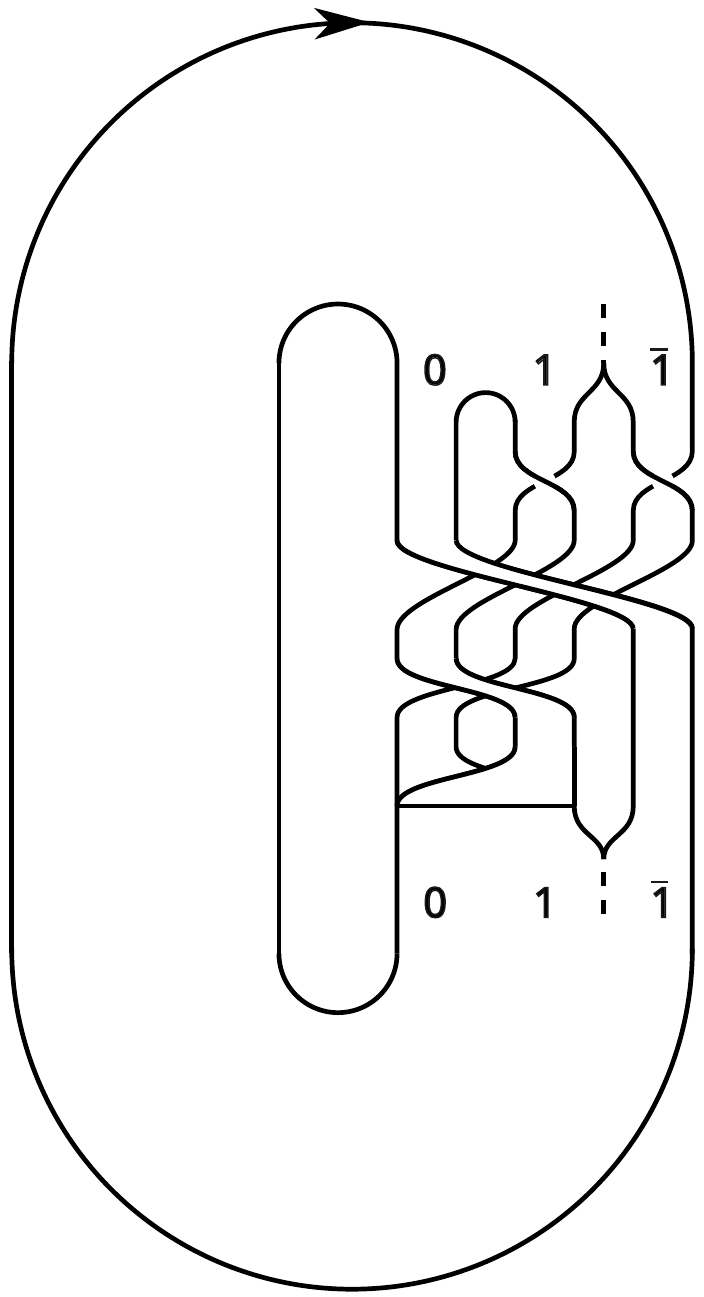}
    \put(-120,140){\large \textbf{B}}
    \\\medskip
    \qquad \qquad \includegraphics[width = .5\textwidth]{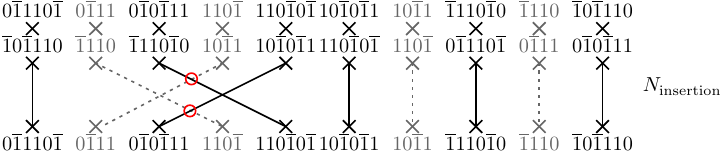}
    \put(-220,30){\large \textbf{C}} \hfill \null
    \caption{
    \textbf{Topological characterisation of the Rössler attractor considering three distinct branches.}
    A. First return map of a Poincaré section with two components, separating out the additional decreasing branch. From branch $0$, only transition to branch $\overline{1}$ are possible.
    B. Template with three branches including a separation between the two components. The separation between $1$ and $\overline{1}$ is virtual as it is underlined by dashed line. \eqref{eq:rossler_trois_bandes} is the linking matrix describing this template.
    C. Linking graph used to theoretically compute linking numbers between orbits $N_\text{insertion}=2$ \eqref{eq:n_insertion}.}
    \label{fig:rossler_01c_appli_orbites}
\end{figure}

Additional branches in first return maps can be observed, but their impact on the topological characterisation is not clear (see \cite{Aguirre_2008, Abadie_2025} where extra-branches are not considered and reciprocally \cite{Letellier_2001, Kamdoum_Tamba_2016} where they are considered). 
These extra-branches must be distinguished from foliated structure exhibited by Lorenz system \cite{lorenz1963deterministic} and Chen system \cite{chen1999yet} where the foliations are the result of a tearing mechanism \cite{Rosalie2016}.
As a first glimpse on the system \eqref{eq:economic} solution, it appears that several folding might occur (Fig.~\ref{fig:economic_attra}) without tearing mechanism.
The purpose of this subsection if to clarify the meaning of an additional branch in first return map where only stretching and folding mechanisms occurs (see template Fig.~\ref{fig:rossler}G).
A first return map is obtained with an extra branch (Fig.~\ref{fig:rossler_01ab_appli_orbites}A) using the Rössler system \eqref{eq:rossler} with the following Poincaré section:
\begin{equation}
  \mathcal{Q} = \left\{ (y_n,-z_n) \in \mathbb{R}^2\ |
  -x_n = -1, \ -\dot{x}_n>0\right\} 
  \label{eq:rossler_01b_section_gamma}
\end{equation}
that is represented in Fig.~\ref{fig:rossler_01b}.
From this first return map, a variable $\gamma_n$ is build to represent the distance from the inside to the outside of the attractor as it has been done for $\rho_n$ \eqref{eq:rossler_01_section_rho} :
\begin{equation}
    \gamma_n = (y_n-0.229)/(1.7634-0.229) .
\end{equation}
As illustrated by Fig.~\ref{fig:rossler_01b}B, such a construction implies the loss of bijection from the Poincaré section $\mathcal{Q}$ to $\gamma_n$.
Facing this extra branch in first return map, one can be tempted to consider that there are three branches in the template with an encoding that could be $0$, $1$ and $\overline{1}$.
Instead of trying to study directly the extra-branch impact from the first return map Fig.~\ref{fig:rossler_01ab_appli_orbites}B, we proposed to build a bijective representation of $\mathcal{Q}$ using $\eta_n$ that is defined as follows:
\begin{equation}
    \eta_n = \left\{
        \begin{aligned}
            &(y_n-0.229)/(1.7634-0.229) \quad &\text{ if } z_n \leq -1.1665 \\
            &1 + (y_n-1.137)/(1.7634-1.137) & \text{ else}
        \end{aligned}
    \right.
\end{equation}
Such a partition of the Poincaré section implies a separation and this is used to characterise chaotic attractor bounded by higher genus torus where multiple components are required to study their topology \cite{tsankov2004topological, Rosalie2014}. 
The first return map associated to this representation of the Poincaré section $\mathcal{Q}$ with $\eta_n$ is represented Fig.~\ref{fig:rossler_01c_appli_orbites}A.

Consequently, $\eta_n$ has values between 0 and 1 giving the increasing and decreasing branches encoded with 0 and 1 from $\gamma_n$ and values of $\eta_n$ between 1 and 2 corresponds to the values of the additional decreasing branch from $\gamma_n$ and is encoded with $\overline{1}$.
The first return map (Fig.~\ref{fig:rossler_01c_appli_orbites}A) associated to such a Poincaré section clearly split the three branches and indicates that there are only three transitions between the components :
\begin{itemize}
    \item From branch $0$, the next periodic point will be in branch $\overline{1}$.
    \item From branch $\overline{1}$, the next periodic point will be in branches $0$ or $1$.
    \item From branch $1$, the next periodic point will be in branches $0$ or $1$.
\end{itemize}

The hypothesis of separating the two decreasing branches leads us to a template (Fig.~\ref{fig:rossler_01c_appli_orbites}B) described by the linking matrix $R_3$:
\begin{equation}
\label{eq:rossler_trois_bandes}
  R_3 =
  \begin{matrix}
  0 \\ 1 \\ \overline{1}
  \end{matrix} \left.\begin{bmatrix}
    0 & -1 & -1 \\
    -1 & -1 & -1 \\
    -1 & -1 & -1 \\
    \end{bmatrix}\!\!\!\right] \;.
\end{equation}
with the following structure (Fig.~\ref{fig:rossler_01c_appli_orbites}B).
This template is also validated using numerical and theoretical linking numbers. This calculation for two periodic orbits of period 4 and 6 is:
\begin{equation}
    \label{eq:n_insertion}
    \begin{aligned}
      lk(10\overline{1}1,10\overline{1}0\overline{1}1) 
    &=\frac{1}{2}\left[ 2R_{0,0} + 6R_{0,1} + 4R_{0,\overline{1}} +
    4R_{1,1} + 6R_{1,\overline{1}} + 2R_{\overline{1},\overline{1}} + N_\text{insertion}\right]\\ 
      &= \frac{1}{2}(0-6-4-4-6-2+2)=-10 \;,
 \end{aligned}
\end{equation}
including the insertion number computed theoretically (Fig.~\ref{fig:rossler_01c_appli_orbites}C).

Considering that both torsions of branch $1$ and $\overline{1}$ are equal to $-1$ and that they permute negatively once, they can be regrouped in a unique negative torsion. 
Consequently, the template with three branches (Fig.~\ref{fig:rossler_01c_appli_orbites}B) is topologically equivalent to the Rössler template with two branches (Fig.~\ref{fig:rossler}G).
Therefore, splitting the first return map into three branches produces the same result as considering only two branches and reordering the periodic points based on their implication number that corresponds to the progression from one side to the other on the first return map.
As the Poincaré section leading to this calculation are equivalents, \eqref{eq:rossler_01_section_rho} and \eqref{eq:rossler_01b_section_gamma}, the additional branch appears because of the positioning of Poincaré section \eqref{eq:rossler_01b_section_gamma} in the chaotic mechanism where stretching and folding are occurring and before the squeezing mechanism.
Consequently, even if the Poincaré section is not ideally positioned to analyse the topology of a chaotic attractor, the shape of its first return map synthesises the chaotic mechanism without the need of considering additional branches as being new interesting branches from the topological perspective.

\section{Template and its implication}
\label{sec:template}

\subsection{Template of the model}

% TODO @Alexandre Mettre ici les orbites, leur encodage, les nombres d'enlacement et le gabarit avec toutes les explications.
\begin{figure}[hbt]
    \centering
    \includegraphics[width = .45\textwidth]{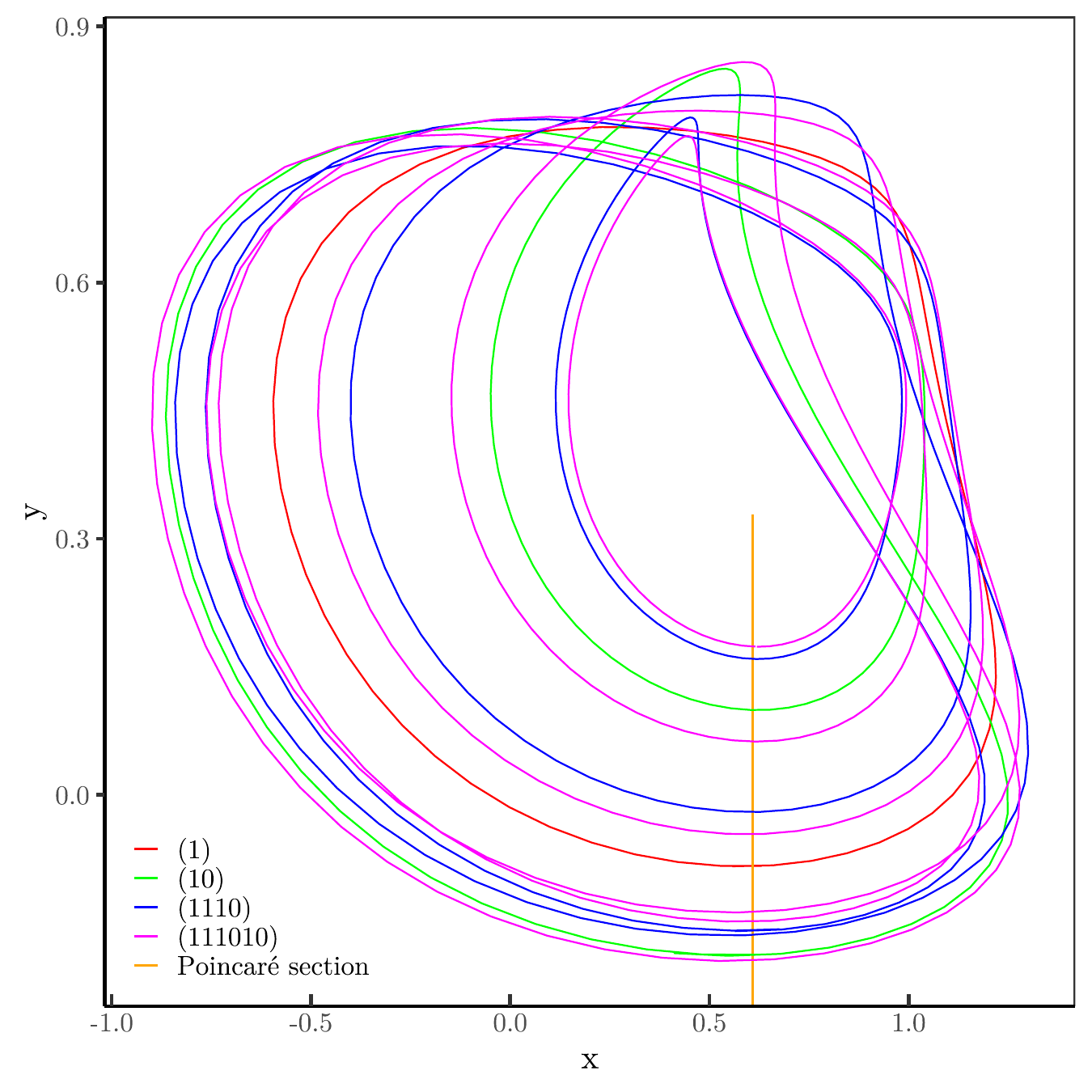}
    \put(-180,160){\large \textbf{A}}
    \quad
    \includegraphics[width = .45\textwidth]{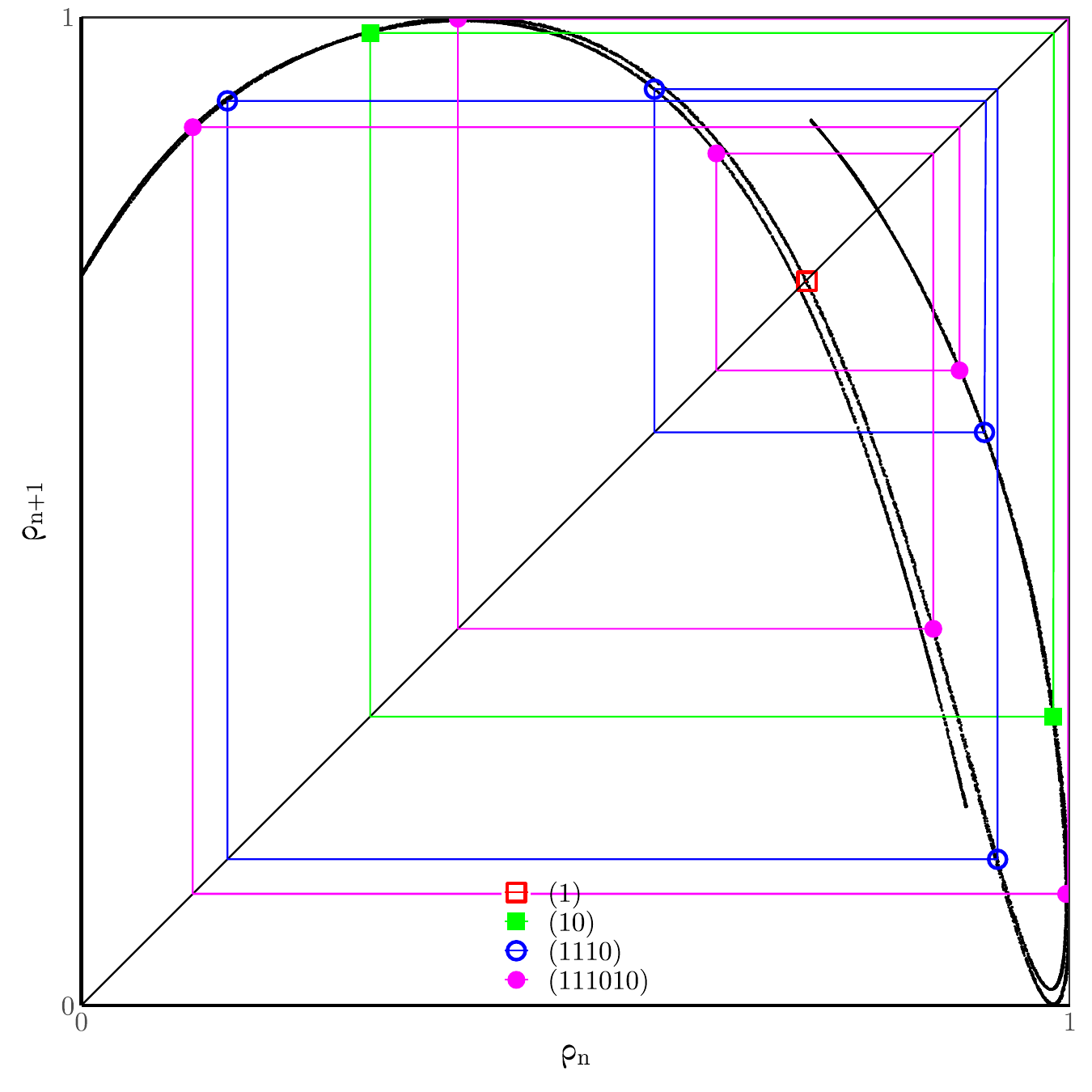}
    \put(-180,160){\large \textbf{B}}
    \caption{ (\textbf{A}) A set of 4 orbits of the attractor $\mathcal{E}$. (\textbf{B}) The first return map associated to $\mathcal{P}$. The 4 orbits are plotted on it. The coordinates are normalized between 0 and 1 \eqref{eq:rho_n}}.
    \label{fig:orbites}
\end{figure}

\begin{table}[htb]
    \centering
    \caption{Numerical linking numbers between pairs of orbits extracted from the chaotic attractor $\mathcal{E}$.}
    \label{table:linking_num}
    \centering
    \begin{tabular}{ccccc}
        \\[-0.3cm]
        \hline \hline
        \\[-0.3cm]
        & (1) & (10) & (1110)  \\[0.1cm]
        \hline
        \\[-0.3cm]
        (10) & -1 & \\[0.1cm]
        (1110) & -2 & -3 & \\[0.1cm]
        (111010) & -3 & -5 & -10 \\[0.1cm]
        \hline \hline
    \end{tabular}
\end{table}

Further to the results of previous section, the topological method is applied directly without considering the layers of $\mathcal{E}$.
Thus, $\rho_n$ is a discrete value detailing successive passage of the flow through the Poincaré section, allowing the drawing of a first-return map that describes the dynamic of the flow in one dimension.
\begin{equation}
\label{eq:rho_n}
\rho_n = \frac{-y-\min(y)}{\max(y)-\min(y)}
\end{equation}

The unstable periodic orbits (Fig.~\ref{fig:orbites}.A) are extracted from the attractor $\mathcal{E}$. 
The first-return map (Fig.~\ref{fig:orbites}.B) is subsequently plotted along with the orbits. 
Based on the work previously detailed (Sec.~\ref{ssec:rossler}, only two branches are considered that leads to the naming convention, $0$ and $1$, used in Fig.~\ref{fig:orbites}, Tab.~\ref{table:linking_num} and Tab.~\ref{table:linking_theo}. 
The symbol $0$ is for the increasing branch and the symbol $1$ for the decreasing branch.
The linking numbers are computed using a script written in Python \cite{uribarri}, giving numbers identical to the one obtained by an analysis of the Rössler system (Fig.~\ref{fig:rossler}). 
The same template and its associated linking matrix are considered:
\begin{equation*}
    L = \left.\left[ \begin{matrix}
    0 & -1 \\ -1 & -1
    \end{matrix} \right]\!\!\right].
    \label{eq:model_template}
\end{equation*}
The computation of the theoretical linking numbers is then done (and detailed in ~\ref{table:linking_theo}) using the following formula: 
\begin{equation*}
    \displaystyle lk(\mathcal{O},\mathcal{O}') = \frac{1}{2}\times\left(\sum_{i=1}^{k}\sum_{j=1}^{l} L_{ \mathcal{O}_i,\mathcal{O}'_j} + N_{joining}(\mathcal{O},\mathcal{O}')\right)
\end{equation*}
As the theoretical linking numbers corresponds to numerical ones, template is validated (Fig.~\ref{fig:rossler}G). 
This template evolving clockwise is made of a continuous part from bottom right to top right.
The remaining part is composed of the chaotic mechanism made with two branches: $0$ that has no torsion and $1$ with a negative torsion permuting with the other branch, stretching and reversing the order of orbits in the flow until both branches are squeezed together.

\begin{table}[tb]
    \centering
    \caption{Theoretical linking numbers between pairs of orbits extracted from the chaotic attractor $\mathcal{E}$.}
    \label{table:linking_theo}
        \resizebox{.6\textwidth}{!}{
        \begin{tabular}{c}
            \\[-0.3cm]
            \hline \hline
            \\[-0.3cm]
            $\begin{aligned}
                lk(1,1110)= & \frac{1}{2}\times[ 3L_{1,1} + L_{1,0} + N_{joining}] \\
                = & \frac{1}{2} \times (-3-1+0)=-2 
            \end{aligned}$ \\[1cm]
            \hline \\
            $\begin{aligned}
                lk(10,1110)= & \frac{1}{2}\times[ 3L_{1,1} + L_{1,0} + 3L_{0,1}  + L_{0,0} + N_{joining}] \\
                = & \frac{1}{2} \times (-3-1-3+0+1)=-3 
            \end{aligned}$ \\[1cm]
            \hline \\
            $\begin{aligned}
                lk(1110,111010)= & \frac{1}{2}\times[ 3\times4L_{1,1} + 3\times2L_{1,0} + 4L_{0,1} + 2L_{0,0} + N_{joining}] \\
                = & \frac{1}{2} \times (-12-6-4+0+2)=-10 
            \end{aligned}$ \\[1cm]
            \hline \hline
        \end{tabular}
        }
\end{table}

\begin{figure}[tb]
    \centering
    \includegraphics{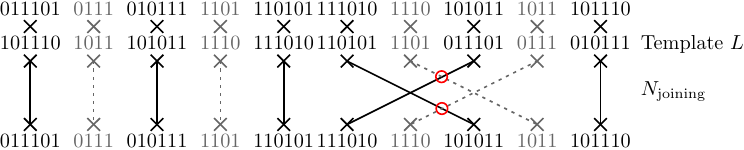}
    \caption{Theoretical computation of $N_{joining}$ for $lk(1110,111010)$. Each positive crossing is counted: $N_{joining}=2$.}
    \label{fig:n_insertion}
\end{figure}

\subsection{Interpretation and discussion}

The main interest of the template representation is that it reproduces all properties of chaotic solutions. 
First, the solution is bounded in the template. Secondly, the evolution is not time-dependent because it is built on topological invariants. 
Finally, the last part of the chaotic mechanism with stretching and squeezing illustrates the sensitivity to initial conditions as well as the nondeterministic properties by allowing all trajectories to be redistributed in every branch for the next topological period.
The topological structure of the attractor $\mathcal{E}$ indicates where the mechanisms leading to chaos occur: stretching to split closed trajectories and folding and squeezing to mix them while maintaining a bounded evolution of the trajectories.
This mechanism is detailed theoretically Fig.~\ref{fig:rossler}G where the periodic orbits are embedded in the template of $\mathcal{R}$ that is equal to the template of $\mathcal{E}$. 
Two branches are distinguished where the left part contains no torsion, and the external branch has a negative torsion.
Orbits of the attractor $\mathcal{E}$ (Fig.~\ref{fig:orbites}A) are spatially closed to the theoretical representation of the template except that there is an additional branch.
The latter indicates that for a couple $(x, y)$ (unemployment and inflation) two distinct values of $z$ (exchange rate) (Fig.~\ref{fig:poincaresec}B) have non equivalent values at the end of the chaotic mechanism in the template.
The stretching, folding and squeezing occur in the chaotic mechanism of the template while a part of the attractor will be squeezed later in the chaotic mechanism after a revolution.
The trajectories on the right of the Poincaré section of Fig.~\ref{fig:orbites}A reproduces transformations due to the chaotic mechanism of the attractor.
As mentioned before, for $\mathcal{E}$, there is no place to position the Poincaré section without having this additional branch. 
This branch is visible all along the flow with two layers.

From an economic point of view, when unemployment decrease, inflation tends to accelerate with possibly two distinct values of exchange rates due to this additional branch.
This parallel evolution of trajectories (which corresponds to the theoretical form of the Philips curve) is maintained until stretching and folding occur to mix trajectories, that is, when both $x$ and $y$, respectively, unemployment and inflation are high (Fig.~\ref{fig:orbites}A).  
This phenomenon, known as stagflation, precede the chaotic mechanism in the phase space. 
In this area, there are several possibilities for closed trajectories, it includes both scenarios where two closed trajectories will diverge to follow distinct paths or, two distinct trajectories could also be closed after the chaotic mechanism because of the squeezing mechanism. 
Another possibility is that two distinct trajectories could exchange their relative positions because at the end of the template, the branches are stretched and squeezed to the branch line.

Stagflation directly contradicts the conventional Phillips curve framework. It has necessitated a re-evaluation of the factors driving inflation and unemployment, prompting economists to consider alternative explanations (supply-side shocks, role of inflationary expectations). Beyond these dimensions that are not addressed in the paper, our empirical approach suggests, on the one hand, that the exchange rate intervenes in this interrelationship between inflation and unemployment, and on the other hand, that stagflationary evolutions (and the resulting monetary policy) can lead to chaotic mechanism.

\section{Conclusion}

A model defined by a three-differential equations system has been obtained, that will hopefully pave the way and bring up interest to the study of non-linear, chaotic dynamics and global modelling based on economic data.
From macroeconomic data time series, a model has been generated using a global modelling tool and its solution has been subsequently topologically characterised.
We proved that the presence of a supplementary branch does not raise any issue in the topological analysis of the attractor $\mathcal{E}$, which is topologically equivalent to that of the classical Rössler attractor.
In terms of economic analysis, our exploratory approach suggests the exchange rate plays a role in the inflation-unemployment relationship: it could lead to very different dynamics that may be considered in theoretical modelling. More important, our tool identified a chaotic mechanism specifically during periods of stagflation -and not elsewhere- thereby indicating where trajectories intertwine. Stagflation, characterised by the simultaneous occurrence of high inflation and economic stagnation, presents a conundrum that challenges traditional macroeconomic models, particularly those based on the Phillips curve, which posits an inverse relationship between inflation and unemployment. In this context, stagflation could be viewed as a transient regime or a shift to another economic state, driven by factors that introduce non-linearities and sensitivity to initial conditions, characteristics of chaotic systems. 
For future works, we plan to realise bifurcation diagrams of the differential equation system to fully understand the route to chaos by varying parameters and contribute to a better understanding of the rise of chaotic mechanisms in the model and its potential position in the economical phase space.

\subsection*{Acknowledgements}

This work has benefited from the support of the ‘Fédération de Recherche Energie et Environnement’ (FREE 2043 CNRS-UPVD) awarded to both Vincent LIGNON and Martin ROSALIE.
The authors thanks Sylvain MANGIAROTTI for helpful comments and remarks.

\subsection*{Conflict of Interest Statement}

The authors have no conflicts to disclose.

\subsection*{Data Availability Statement}

The data that support the findings of this study are available from the corresponding author upon reasonable request.

\printbibliography

\begin{figure*}[p]
    \caption{Graphical abstract}
\includegraphics[width = \textwidth]{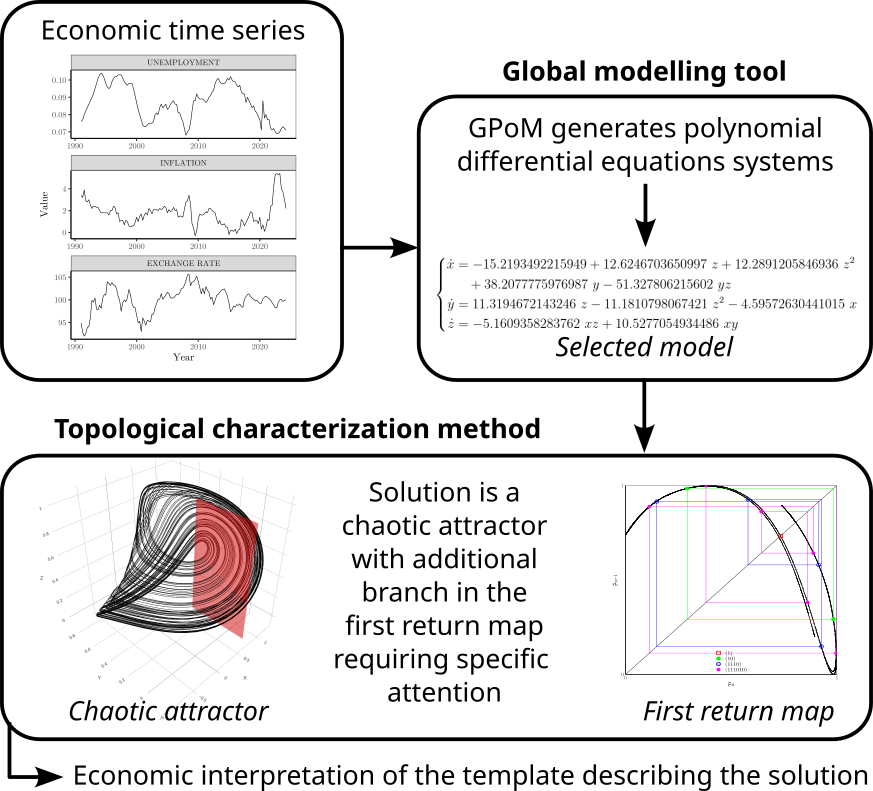}
\end{figure*}

\end{document}